\documentclass[submission,copyright,creativecommons]{eptcs}

\usepackage{amsmath,amssymb,amsfonts}
\usepackage{graphicx}
\usepackage{textcomp}
\usepackage[table]{xcolor}
\usepackage[linesnumbered,ruled,noend]{algorithm2e}
\usepackage[utf8]{inputenc}

\usepackage[rflt]{floatflt}
\usepackage{xspace}
\usepackage{changes}
\usepackage[caption=false]{subfig}

\usepackage{tikz}
\usetikzlibrary{automata,backgrounds,calc,fit,positioning,shapes}
\usepackage{pgfplots}
\usepgfplotslibrary{fillbetween}
\usepackage[overlay,absolute]{textpos}
\usepackage{transparent}
\usepackage{multicol}

\def\papername{\jobname}

\usepackage{multirow}

\g@addto@macro\UrlBreaks{\do\-}
\newcommand{\simulink}{\textsc{Simulink}}
\newcommand{\profeat}{\textsc{ProFeat}}
\newcommand{\python}{\textsc{Python}}
\newcommand{\prism}{\textsc{Prism}}
\newcommand{\storm}{\textsc{Storm}}
\newcommand{\errorpro}{\textsc{OpenErrorPro}}

\newcommand{\nusmv}{\textsc{NuSMV}}
\newcommand{\simpars}{\textsc{SimPars}}

\newcommand{\cB}{\ensuremath{\mathcal{B}}\xspace}

\newcommand{\cD}{\ensuremath{\mathcal{D}}\xspace}

\newcommand{\cP}{\ensuremath{\mathcal{P}}\xspace}

\newcommand{\fD}{\ensuremath{\mathcal{D}}\xspace}

\newcommand{\power}[1]{{\wp}\left(#1\right)}

\newcommand{\Nat}{\mathbb{N}}

\newcommand{\Eval}{\mathit{Eval}}
\newcommand{\Var}{\mathit{Var}}

\newcommand{\Prog}{\mathsf{Prog}}
\newcommand{\reorder}{\mathit{reorder}}
\newcommand{\size}{\mathit{size}}
\newcommand{\cnf}{\mathit{cnf}}

\newcommand{\lsem}{[\![}
\newcommand{\rsem}{]\!]}
\newcommand{\true}{\textsf{true}}
\newcommand{\false}{\textsf{false}}

\providecommand{\event}{MARS 2020} %
\usepackage{breakurl}             %

\title{Iterative Variable Reordering: Taming Huge System Families\thanks{%
	This work is supported by the DFG through
the Collaborative Research Centers CRC 912 (HAEC) and
TRR 248 (see {\footnotesize\url{https://perspicuous-computing.science}}, project ID 389792660),
the Cluster of Excellence EXC 2050/1 (CeTI, project ID 390696704, as part of Germany's Excellence Strategy),
the Research Training Groups QuantLA (GRK 1763) and RoSI (GRK 1907), 
projects JA-1559/5-1, BA-1679/11-1, BA-1679/12-1, and the 5G Lab Germany.}}
\def\titlerunning{Iterative Variable Reordering: Taming Huge System Families}

\author{Clemens Dubslaff
\institute{Institute of Theoretical Computer Science\\
Technische Universität Dresden\\ Dresden, Germany}
\email{clemens.dubslaff@tu-dresden.de}
\and
Andrey Morozov
\institute{Institute of Automation\\
Technische Universität Dresden\\ Dresden, Germany}
\email{andrey.morozov@tu-dresden.de}
\and
Christel Baier
\institute{Institute of Theoretical Computer Science\\
Technische Universität Dresden\\ Dresden, Germany}
\email{christel.baier@tu-dresden.de}
\and
Klaus Janschek
\institute{Institute of Automation\\
Technische Universität Dresden\\ Dresden, Germany}
\email{klaus.janschek@tu-dresden.de}
}
\def\authorrunning{C. Dubslaff, A. Morozov, C. Baier, and K. Janschek}
\begin{document}
\maketitle

\begin{abstract}
For the verification of systems using model-checking techniques, 
symbolic representations based on binary decision diagrams (BDDs) 
often help to tackle the well-known state-space explosion problem. 
Symbolic BDD-based representations have been also shown to be 
successful for the analysis of families of systems that arise, e.g., 
through configurable parameters or following the feature-oriented 
modeling approach. The state space of such system families face an 
additional exponential blowup in the number of parameters or 
features. It is well known that the order of variables 
in ordered BDDs is crucial for the size of the model representation. 
Especially for automatically generated models from real-world systems, 
family models might even be not constructible due to bad variable orders. 
In this paper we describe a technique, called \emph{iterative variable reordering}, 
that can enable the construction of large-scale family models. We exemplify 
feasibility of our approach by means of an aircraft velocity control system 
with redundancy mechanisms modeled in the input language of the probabilistic 
model checker $\prism$. We show that standard reordering and dynamic reordering 
techniques fail to construct the family model due to memory and time constraints, 
respectively, while the new iterative approach succeeds to generate a symbolic 
family model.
\end{abstract} 

\section{Introduction}\label{sec:introduction}
Model checking is an automated technique for the verification of systems~\cite{CGP00,BK08},
applied in many areas of system design where reliability and correctness are key.
The biggest challenge model checking faces is the well-known
\emph{state-explosion problem} that describes the exponential
blow up of states in the number of system variables.
One prominent approach to cope with the state-explosion problem 
in model checking is the use of symbolic methods, such as those based on
binary decision diagrams (BDDs) \cite{BCMcMDH92,McMillan}.
Such symbolic methods are most viable the more redundant or shared behaviors 
are present in the system description, as symbolic methods
can exploit them towards concise representations.
It is hence not surprising that symbolic model-checking techniques
also have been successfully applied for the analysis of system 
families those members share lots of common behaviors.
Most prominently, in feature-oriented system analysis~\cite{Thum14},
a member of the system family is characterized by its basic functionalities
by means of \emph{features}. Features can either be active or inactive, 
possibly leading to system families those size is exponential in the
number of features. In case family members have active features
in common, they likely share the behaviors of these features, 
making symbolic methods effective. 
Despite for feature-oriented systems, also other kinds of configurable 
systems can benefit from symbolic methods as one easily faces an exponential 
blow up in the number of configuration parameters.

There are mainly two approaches to construct and analyze system families. 
The first, called \emph{one-by-one approach}, 
constructs and analyzes the model for each family member separately. 
Differently, within an \emph{all-in-one approach}, a single family model is constructed
and analyzed in a single run. For system families those members share a 
lot of common behaviors, all-in-one approaches can benefit from 
symbolic representations and analysis methods by efficiently
representing common parts. This usually leads to better analysis times
than for one-by-one approaches~\cite{ClassenHSL11,Thum14,DBK15,CDKB18} 
or even enables an analysis~\cite{KBCDDKMM18,DDMBJ19}.\\

In this paper, we focus on all-in-one approaches where the family model 
is symbolically represented using reduced ordered BDDs. Such BDDs are 
rooted directed graphs where inner nodes are labeled by variable names,
each path obeying a given variable order. It is well known that the
size of the BDD significantly depends on the chosen variable 
order~\cite{Bryant86,KBCDDKMM18}.
In real-world applications, models subject to verification are usually
automatically generated such that ad-hoc variable orders are likely to provide
bigger BDD representation than possible with another variable order
or even lead to BDD sizes exceeding memory constraints.
Reordering algorithms such as sifting~\cite{Rud1993,Panda95} help
to find suitable variable orders towards a small symbolic representation
but are not applicable when the model cannot be constructed a priori due
to insufficient memory. A solution is to dynamically reorder BDD variables
\cite{FelYorBra93a} during model construction, which however comes at the 
cost of spending much time on reordering.
For system families where the family model cannot
be constructed due to bad variable orders and dynamic reordering techniques
exceed time constraints, we present a third automated method we call \emph{iterative
variable reordering}. The basic idea is to construct parts of the 
system family, apply variable-reordering techniques 
towards smaller BDD representations, and then successively add family 
members until a suitable variable order
for the whole family model is found. In this way, the shared behaviors 
between family members are step-wise incorporated into the symbolic representation.
We first used this idea in~\cite{DDMBJ19} to enable 
the construction of a family model of aircraft velocity control loops
given in the input language of the prominent probabilistic model 
checker $\prism$~\cite{KNP11}. The model of \cite{DDMBJ19} 
was automatically generated from $\simulink$ code~\cite{aircraft} 
using $\simpars$~\cite{MorJanKru16a}, initially not constructible
using the symbolic engine of $\prism$ based on 
\emph{multi-terminal BDDs}~\cite{CFMcGMcMY93,mtbdds,Hachtel-Som-et-al-ADD97}.
Starting with the most basic variant of the aircraft velocity control loop, 
we step-wise added system variants and applied automated variable reordering capabilities presented in~\cite{KBCDDKMM18} to obtain a suitable
variable order such that the whole family model could be constructed.
While~\cite{DDMBJ19} focused on enabling a reliability analysis of the
aircraft velocity control loop, we approach iterative variable
reordering in a generic fashion in this paper and present an automated
variant of the ad-hoc and handcrafted concepts of~\cite{DDMBJ19}.
Specifically, our contribution is as follows:
\begin{itemize}
	\item we formally specify iterative variable reordering
		along with heuristics,
	\item present an implementation of the algorithm that enables
		the automated application of iterative variable reordering
		on $\prism$ family models, and
	\item evaluate the approach on an even bigger instance of the
		velocity control loop model from \cite{DDMBJ19}, showing
		the effectiveness of our implementation.
\end{itemize}
To the best of our knowledge, we are the first who present a variable 
reordering technique that is specifically tailored for family models.

\paragraph{Outline.}
In Section~\ref{sec:preliminaries},
we provide foundations on abstract programs that formalize our models,
revisit BDDs and variable reordering as well as $\prism$ programs.
The generic algorithm of iterative variable reordering is subject
of Section~\ref{sec:reordering}. In Section~\ref{sec:model} we
briefly recall redundancy systems and the aircraft velocity control loop
model we introduced in~\cite{DDMBJ19} which we will use in
Section~\ref{sec:evaluation} as example for evaluating our implementation.
We close the paper with some concluding remarks in Section~\ref{sec:conclusion}.

\section{Foundations}\label{sec:preliminaries}
For a given set $X$ we denote by $\power{X}$ its power set, i.e., the set of all
subsets of $X$.
We call a total order $(D,<_D)$ on a finite set $D$ an \emph{ordered domain}.
For a finite set of variables $\Var$ we assign to each variable
an ordered domain from a set $\fD$ by a function $d\colon \Var\rightarrow \fD$.
A variable evaluation over $\Var$ w.r.t. $d$ is a function $\eta$ that 
assigns to each variable $v\in\Var$ a value $\eta(v)\in d(v)$.
We denote by $\Eval_d(\Var)$ the set of variable evaluations over $\Var$ w.r.t. $d$.
In the following, we assume the function $d$ to be fixed for a given set of variables $\Var$
and briefly write, e.g., $\Eval(\Var)$ for the set of evaluations over $\Var$.

We consider \emph{programs} $\Prog=(\Var, C, \iota)$ where 
$\Var$ is the finite set of variables on which the program is defined, 
$C$ is a set of commands, and $\iota$ is an expression specifying a set of
initial variable evaluations that we denote by $\lsem\iota\rsem \subseteq \Eval(\Var)$.
Sometimes we use sloppy notations and replace $\iota$ by
a single evaluation $\eta\in\Eval(\Var)$ to represent the 
initial variable evaluation explicitly.
Intuitively, a program defines a state-based semantics where states 
are variable evaluations in $\Eval(\Var)$ and commands specify how to
switch to from one variable evaluation to another.
Every program specifies a family of systems through the initial variable 
evaluations, i.e., we interpret each initial variable evaluation as
entry point for a system variant. This kind of interpretation is backed
by well-known concepts from feature-oriented and family-based system modeling
and analysis~\cite{ClassenHSL11,Thum14,DBK15,CDKB18}.

We intentionally defined programs in a rather abstract way, as the
concepts presented in this paper are applicable to a wide range of 
system families described by various specification formalisms.
Thus, we also leave out to specify a concrete semantics of programs.

\subsection{Symbolic Representations Through Binary Decision Diagrams}
Binary decision diagrams (BDDs) \cite{Lee59,Akers78} were mainly developed 
as a universal data structure for Boolean functions, i.e., a BDD over 
a set of Boolean variables $\Var$ describes a function 
$f\colon \power{\Var} \rightarrow \{\true,\false\}$ that maps a set of variables
assumed to be $\true$ to either $\true$ or $\false$.
Technically, a BDD over $\Var$ is a graph that is rooted, directed, and acyclic and where 
each node is either terminal and represents a result, i.e., either $\true$ or $\false$,
or it is a decision node. Each decision node is labeled by a Boolean variable 
from $\Var$ and has exactly two successor nodes: 
the $0$-successor node, which stands for assigning $\false$ to the respective Boolean variable, 
and the $1$-successor node, standing for a $\true$ assignment, respectively.
We refer to the number of nodes in a BDD $\cB$ as \emph{size}, denoted 
by $\size(\cB)$.
In the context of model checking, BDDs are used to represent the 
characteristic function of the transition relation for state-based models.

\paragraph{Reduced Ordered BDDs.}
An \emph{ordered BDD} \cite{Bryant86} is a BDD over a total order $(\Var,\prec)$
on Boolean variables where along all paths from the root to a terminal node 
the order $\prec$ is respected. 
The graph structure of an ordered BDD for a Boolean function $f$ arises from a binary 
decision tree for $f$ using the given variable ordering. By merging isomorphic 
subgraphs, eliminating terminal nodes with the same value, and removing any node whose 
two successors are isomorphic we obtain \emph{reduced ordered BDDs},
which we simply abbreviate as \emph{BDDs} in the following. 
In such BDDs, every two nodes represent different Boolean functions. 
Removing redundancies from a decision tree reveals the potential of BDDs 
for compactly representing Boolean functions. When BDDs are used to represent 
family models such as programs defined above, behavior shared between family members
could lead to concise BDD representations and analysis of the family.
For further details on BDDs, we refer to standard textbooks such 
as~\cite{Somenzi99binarydecision,Wegener00}. 

\paragraph{Variable Reordering.}
The size of a BDD crucially depends on the given variable ordering~\cite{Bryant86}. 
In fact, there are Boolean functions that can be represented by BDDs of
linear and exponential size depending on the chosen variable order. 
Reordering algorithms such as sifting~\cite{Rud1993,Panda95}, 
can be applied to improve the size BDD for a given Boolean function.
In case BDDs are used as a representation for state-based models, reordering techniques
are applied when the model is constructed. However, when the
constructed model is used for BDD-based verification, dynamic
reordering during the verification process can also be applied, possibly
avoiding bad variable orders in those BDDs that represent intermediate results.
A good heuristic for variable orderings is to first decide variables 
which are ``most-influential'', i.e., changed at the beginning of an execution
of the modeled system and influence the systems
behavior significantly~\cite{MalWanBraSan1988}. 

\paragraph{BDDs over Program Variables.}
In our setting, we use BDDs as representation for state-based models
of programs, which are defined on possibly non-Boolean variables.
However, every element of a finite domain can be represented by a
bit vector. Given a finite domain $D$, we enumerate the domain 
by a function $\#\colon D \rightarrow \{0, \ldots, |D|-1\}$. Then for
an element $x\in D$ we use the bit representation of $\#(x)$ with
length $k=\lceil\log (|D|-1)\rceil$, i.e., introduce $k$ fresh
Boolean variables to represent elements of the domain $D$.
In this way, the interpretation of BDDs on Boolean variables
can easily be lifted to BDDs over variables $\Var$ of programs.
We hence might refer to variables and functions in this general
meaning rather than referring to Boolean variables and Boolean 
functions only. When reordering program variables, we assume 
that the bit-wise order of the variables is maintained (i.e., 
bits are not \emph{exploded}, see~\cite{KBCDDKMM18}).

\subsection{$\prism$ Programs}
We introduced programs in a generic fashion as tuples $\Prog=(\Var,C,\iota)$.
For our case study, we rely on an instance of such programs given
by the input language of the probabilistic model checker $\prism$~\cite{KNP11}.
In $\prism$, variables of $\Var$ are assigned to bounded intervals of integers 
with the standard total order on integers as ordered domain. 
Possible transitions between states, i.e., variable evaluations,
are given by \emph{guarded commands}~\cite{Dijkstra1975}
collected in $C$. A command has the form
\begin{center}
  $\texttt{[action] guard} \enskip \rightarrow \enskip
  \texttt{p}\textsubscript{1}\texttt{:update}\textsubscript{1} \texttt{ + }
  \texttt{p}\textsubscript{2}\texttt{:update}\textsubscript{2} \texttt{ + }
  \ldots  \texttt{ + }
  \texttt{p}\textsubscript{n}\texttt{:update}\textsubscript{n}\enskip.$
\end{center}
Here, \texttt{guard} is a Boolean expression over arithmetic constraints on variable
evaluations, e.g., $\mathtt{(x=1)}\wedge\mathtt{(y\leq5)}$ for $\texttt{x},\texttt{y}\in\Var$
is fulfilled in every state where variable \texttt{x} has value \texttt{1} and
\texttt{y} has a value from the domain $d(\texttt{y})$ of \texttt{y} that is
smaller or equal than \texttt{5}.
In case the guard evaluates to \texttt{true} in some state, the command is
enabled, leading to a transition into a successor state by updating variables 
according to \emph{updates}. An update describes how variables change depending
on the current variable evaluation, e.g., \texttt{x'=1+y} changes the value
of \texttt{x} to the increment of the value of \texttt{y}.
Each update \texttt{update}\textsubscript{i} is chosen with probability 
\texttt{p}\textsubscript{i} for $i\in\{1,\ldots,n\}$.
That is, in every state fulfilling \texttt{guard} the evaluations of
the expressions 
\texttt{p}\textsubscript{1}, \texttt{p}\textsubscript{2}, \ldots,
\texttt{p}\textsubscript{n} constitute a probability distribution,
i.e., must sum up to 1. 
$\prism$ programs describe stochastic state-based models such as
Markov decision processes (MDPs) or discrete Markov chains (DTMCs).
For further details on such models, we refer to standard textbooks
such as \cite{Puterman:book,BK08}.

\paragraph{Family Models in $\prism$.}
In $\prism$, initial variable evaluations describe the entry points for
system variants in families, specified through a guard encapsulated
in an \texttt{init} block. Guards are, as for commands, Boolean expressions
over arithmetic constraints on variable evaluations. All variable evaluations
that fulfill the guard in the \texttt{init} block will be considered as
initial states. For instance, when considering variables
$\Var=\{\texttt{x},\texttt{y}\}$ with domains $d(\texttt{x})=\{0,1\}$ and
$d(\texttt{y})=\{4,5,6\}$, the following \texttt{init} block specifies
two initial states $\eta_1$ and $\eta_2$ where $\eta_1(\texttt{x})=\eta_2(\texttt{x})=1$,
$\eta_1(\texttt{y})=4$, and $\eta_2(\texttt{y})=5$:
\begin{center}$
	\texttt{init\enskip (x=1) }\wedge\texttt{ (y}\leq\texttt{5) \enskip endinit}\enskip.
$\end{center}
The described family model would have two members, one system starting in $\eta_1$
and one starting in $\eta_2$.
In this paper, we also consider \texttt{init} blocks in conjunctive normal
form (CNF) over single variable evaluations. 
The above block could, e.g., be represented in CNF by
\begin{center}$
	\texttt{init\enskip (x=1) }\wedge\texttt{ (y=4)}\vee\texttt{(y=5) \enskip endinit}\enskip.
$\end{center}

\paragraph{Variable Ordering in $\prism$.}
For its symbolic engines, $\prism$ uses algorithms
that rely on Multi-terminal binary decision 
diagrams~\cite{CFMcGMcMY93,mtbdds,Hachtel-Som-et-al-ADD97} (MTBDDs). 
MTBDDs extend BDDs by allowing for terminal nodes with real 
values rather than $\true$ and $\false$. 
In $\prism$ they are used as data structure for representing 
the transition probability matrix of stochastic models. 
To this end, the performance and memory consumption of an analysis 
by $\prism$ crucially depends on the variable order inside the MTBDD
that represents the model. 
In $\prism$, the variable order coincides with the order of 
variable declarations in the program, enabling reordering
methods directly on the source-code level\footnote{Recall
that we considered commands to be unordered in general programs.
We hence always provide a variable order explicitly for $\prism$ programs.}.
For family models in $\prism$, 
\cite{DBK15} showed that a suitable variable order could significantly 
improve the analysis speed using a hand-crafted approach
for tuning variable orders. In \cite{KBCDDKMM18}
automated reordering techniques for $\prism$ models 
have been presented, also showing that the size of the symbolic
representation of the feature-oriented family model of~\cite{DBK15} 
could be reduced even further and yield analysis speed ups.

\section{Iterative Variable Reordering}\label{sec:reordering}

For this section, let us fix a program $\Prog=(\Var,C,\iota)$ on which
we explain our approach of \emph{iterative variable reordering}.
We assume $\lsem\Prog\rsem$ to denote the state-based semantics 
of the program $\Prog$. Furthermore, we assume that there is a 
uniquely defined symbolic representation 
$\lsem\Prog\rsem_\pi$ of $\lsem\Prog\rsem$ that depends
on a \emph{variable order} $\pi=(\Var,\prec)$, i.e., a
total order on the set of variables $\Var$ the program $\Prog$ is
defined on. For instance, when $\Prog$ is a $\prism$ program, 
$\lsem\Prog\rsem_\pi$ could be the MTBDD representation of the
program's semantics in terms of an MDP or DTMC w.r.t. the
variable order $\pi$.

Recall that the purpose of our reordering approach is to enable the
construction of a symbolic representation $\lsem\Prog\rsem_\pi$ of 
system families that are not completely constructible ad-hoc from a given $\pi$.
In this paper, ``constructability'' is understood with respect to
given memory and time constraints.
The presented method can be only effective in case there is a 
variable order $\rho$ such that the complete family model 
$\lsem\Prog\rsem_\rho$ is constructible. Note that even though,
our presented method does not ensure to yield a constructible
family model representation $\lsem\Prog\rsem_\rho$. This is mainly
due to intermediate model representations that might be not constructible.

~\\As prerequisites to our presented method, we rely on the following assumptions:
\begin{enumerate}
	\item[(1)] We have given an initial variable evaluation $\eta\in\lsem\iota\rsem$
		and a variable ordering $\pi$ where $\lsem(\Var,C,\eta)\rsem_\pi$
		is constructible.
	\item[(2)] We have a reordering method $\reorder(\cdot)$ that accepts a
		symbolic representation $\lsem\cP\rsem_\alpha$ of a program $\cP$ and a 
		variable order $\alpha$ and returns a variable order $\beta$ such
		that $\size\big(\lsem\cP\rsem_\alpha\big)\geq\size\big(\lsem\cP\rsem_\beta\big)$.						
\end{enumerate}

\noindent
The above assumptions intuitively provide a base for the iteration (1) and
a method to iterate (2).

We specify iterative variable reordering as pseudo code
in Algorithm~\ref{ireorder-pseudo}, where for some evaluation domain
$E\colon \Var \rightarrow \power{\cD}$ we define its representation as
conjunctive normal form (CNF) by
\[
	\cnf(E) \enskip = \enskip \bigwedge\nolimits_{v\in\Var}
		\bigvee\nolimits_{x\in E(v)} (v = x)\enskip.
\]
Choosing the CNF representing $E$ is a design decision that could be replaced
by any other equivalent representation by Boolean expressions over arithmetic constraints
on variables and domains.
\begin{algorithm}[t]
	\SetAlgoLined
	\DontPrintSemicolon
	\SetKwInOut{Input}{input}\SetKwInOut{Output}{output}
	\Input{Program $\Prog=(\Var,C,\iota)$; variable order $\pi=(\Var,\prec)$ and initial 
		variable evaluation $\eta\in\lsem\iota\rsem$ as in assumption (1)} 
	\Output{Variable order $\rho=(\Var,\prec')$}
	\BlankLine
	Construct $\cB=\lsem (\Var,C,\eta) \rsem_\pi$\;
	$\rho := \reorder(\cB)$\;
	\ForAll{$v\in\Var$}{
		$E(v) := \{\eta(v)\}$\;
		$G(v) := \{\eta'(v) : \eta'\in\lsem\iota\rsem\}$\;
	}
	\While{$E \neq G$}{
		Pick the $\pi$-minimal $v\in\Var$ where $E(v)\neq G(v)$\;
		$E(v) := E(v) \cup \{\min_{d(v)}(G(v)\setminus E(v))\}$\;
		Construct $\cB=\lsem (\Var,C,\iota\wedge\cnf(E)) \rsem_\rho$\;
		$\rho := \reorder(\cB)$\;
	}     
	\Return $\rho$ \;	
	\caption{Iterative variable reordering}
	\label{ireorder-pseudo}
\end{algorithm}
The algorithm is instantiated with the initial variable order $\pi$
and initial evaluation $\eta$ from which we know by assumption (1) that the 
symbolic representation $\lsem(\Var,C,\eta)\rsem_\pi$ is constructible.
The actual construction is performed in line 1, directly followed by a reordering
step to initialize the variable order $\rho$ (line 2).
Then, the evaluation domain $E$ based on $\eta$ is specified in line 4, while
in line 5 the \emph{goal} evaluation domain $G$ provides the sets of all values a variable
can evaluate to in an initial variable evaluation of $\iota$.
The algorithm then iteratively adds values for variables $v$ to the evaluation domain $E$
that are used in an initial variable evaluation of $\iota$ until the goal evaluation
domain $G$ is reached (line 6-10). This is done by always selecting $v$ as the variable  
that is minimal w.r.t. $\rho$. After constructing $\lsem (\Var,C,\iota\wedge\cnf(E)) \rsem_\rho$,
i.e., the symbolic representation of the program $\Prog$ with the additional option
evaluating $v$, the symbolic representation is reordered using $\reorder(\cdot)$
and $\rho$ is updated to a new variable order $\beta$ for which 
\[
	\size\big(\lsem (\Var,C,\iota\wedge\cnf(E)) \rsem_\rho\big)
	\geq
	\size\big(\lsem (\Var,C,\iota\wedge\cnf(E)) \rsem_\beta\big)
\]
due to assumption (2).
The algorithm terminates always in case the intermediate symbolic representations
could be constructed (see line 9) as $E$ is strictly increasing, $G$ is finite,
and in each iteration step and for all $v\in\Var$ we have $E(v)\subseteq G(v)$ as an invariant.

\subsection{Optimization Heuristics}
The basic iterative variable reordering algorithm provided by Algorithm~\ref{ireorder-pseudo}
can be extended by various optimization heuristics. Such heuristics
could enable or fasten the process of finding a suitable variable reorder $\rho$ 
that allows for the construction of $\lsem\Prog\rsem_\rho$.
We focus here on the following variations:
\begin{description}
	\item[Step size:] 
		For a given integer $n\in\Nat$ called \emph{step size}, we repeat line 7
		and 8 $n$ times, i.e., increase the evaluation domain $E$ by $n$ elements.
	\item[$\rho$-min/max selection:]
		In line 7, we choose the $\rho$-minimal or $\rho$-maximal $v\in\Var$ instead the $\pi$-minimal.
\end{description}
The ratio behind the step-size heuristics is that programs with many variables might
waste a lot of time with reordering while adding more family members in one iteration
while they could also provide a suitable variable order where the symbolic 
representation is already constructible. 
The two latter heuristics provide a form of adaptivity to the
selection of variables. For instance, when $\pi$ already has most influential variables
at the beginning of the order, adding family members for influential variables could 
lead to symbolic representations not constructible anymore or, the other way around, less
influential variables in front of the variable order could only lead to small changes
in the reordered order. Adding variants to variables usually increases their influence on the
operational behavior. It is likely that then, this variable appears more in front
of the reorder, following the rule of ``most influential'' variables~\cite{MalWanBraSan1988}
in case for BDDs as symbolic representation.
Choosing $\rho$-maximal variables adds variants to those variables that are not as influential,
somehow balancing their effect on the size of the symbolic representation.

\section{Redundancy System Models}\label{sec:model}
To support our implementation and evaluation of the iterative 
variable reordering, this section is devoted to a brief 
summary of the approach of~\cite{DDMBJ19} and the models
investigated in this paper.

\begin{figure*}[hbtp]
	\centering\includegraphics[width=.86\textwidth]{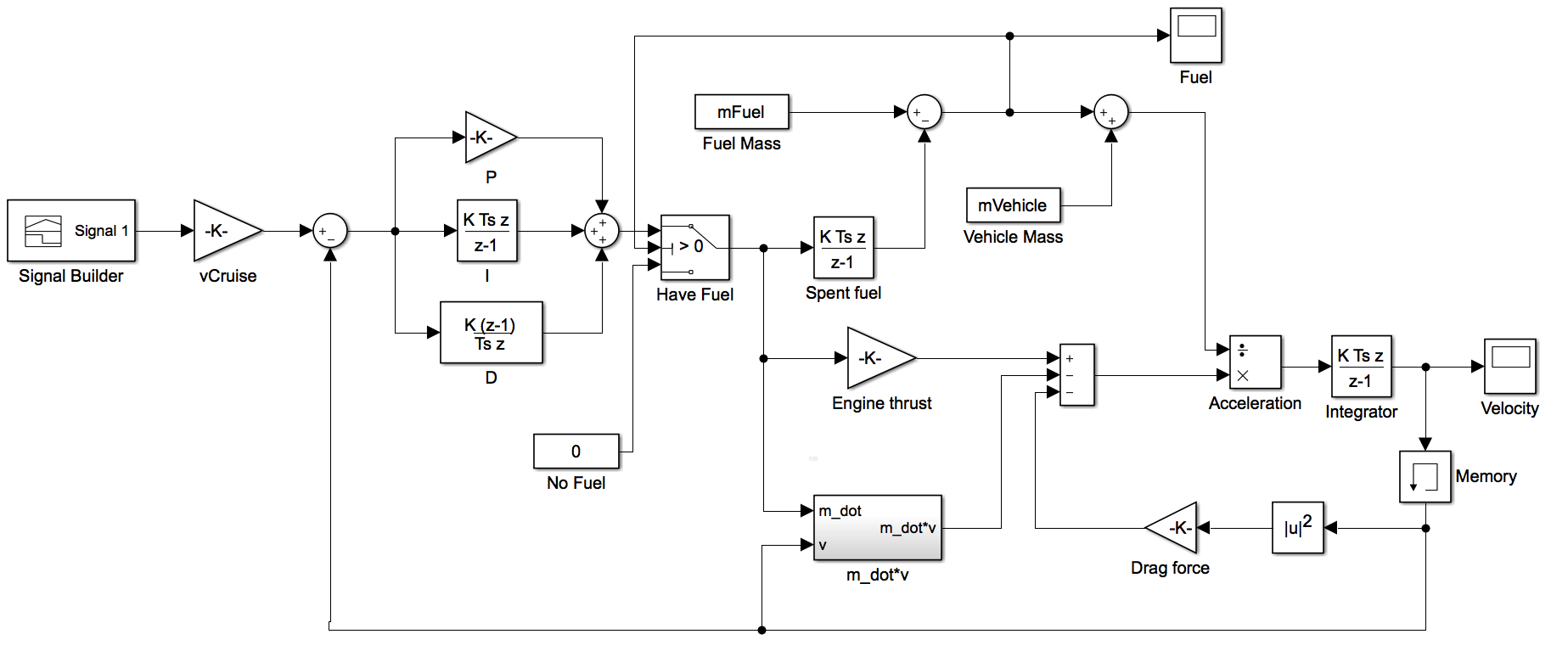}
	\caption{\label{fig:velocity} The $\simulink$ aircraft velocity control loop (VCL) model}
\end{figure*}

\emph{Redundancy systems} are systems where components might be 
replicated to increase fault tolerance of the overall system.
For instance, some component could be protected by triplicating
the component and process the replicas' results through a 
majority voting mechanism towards a single output, implementing
the well-known \emph{triple modular redundancy (TMR)} principle.
Analyzing each instance of the redundancy system easily becomes 
infeasible as the number of possible protection combinations 
grows exponentially in the number of protectable components.
In \cite{DDMBJ19} we proposed to use family-based approaches
for the analysis of redundancy systems, 
motivated by the fact that introducing redundancies in components
also introduces common behaviors. Such common behaviors
are likely to be concisely representable using symbolic techniques.
We illustrated the benefits of our approach by a reliability analysis 
of redundancy systems modeled in $\simulink$~\cite{Simulink},
a widely used framework for model-based system design. 
A $\simulink$ design comprises blocks that describe the
behavior of the system, connected by arrows standing for control and data flow.
An example of a $\simulink$ design is provided in Figure~\ref{fig:velocity},
borrowed from the $\simulink$ example set~\cite{aircraft}.

\subsection{$\simulink$ Models with Redundancy}\label{ssec:simulinkredundancy}
To introduce redundancy mechanisms into $\simulink$ models, we consider
syntactic transformation rules that describe how to obtain protected 
blocks from non-protected ones.
In this paper, we consider the following redundancy mechanisms~\cite{DDMBJ19}:
\begin{description}
	\item[(comparison)] The block is duplicated and both outputs are compared.
		In case their output differs a dedicated failure state is reached. 
		Otherwise, the output is the one of both blocks.
	\item[(voting)] Following the principle of triple modular 
		redundancy principle, the block is triplicated and the 
		output is based on a majority decision.
\end{description}

\begin{figure*}[hbtp]
	\includegraphics[width=\textwidth]{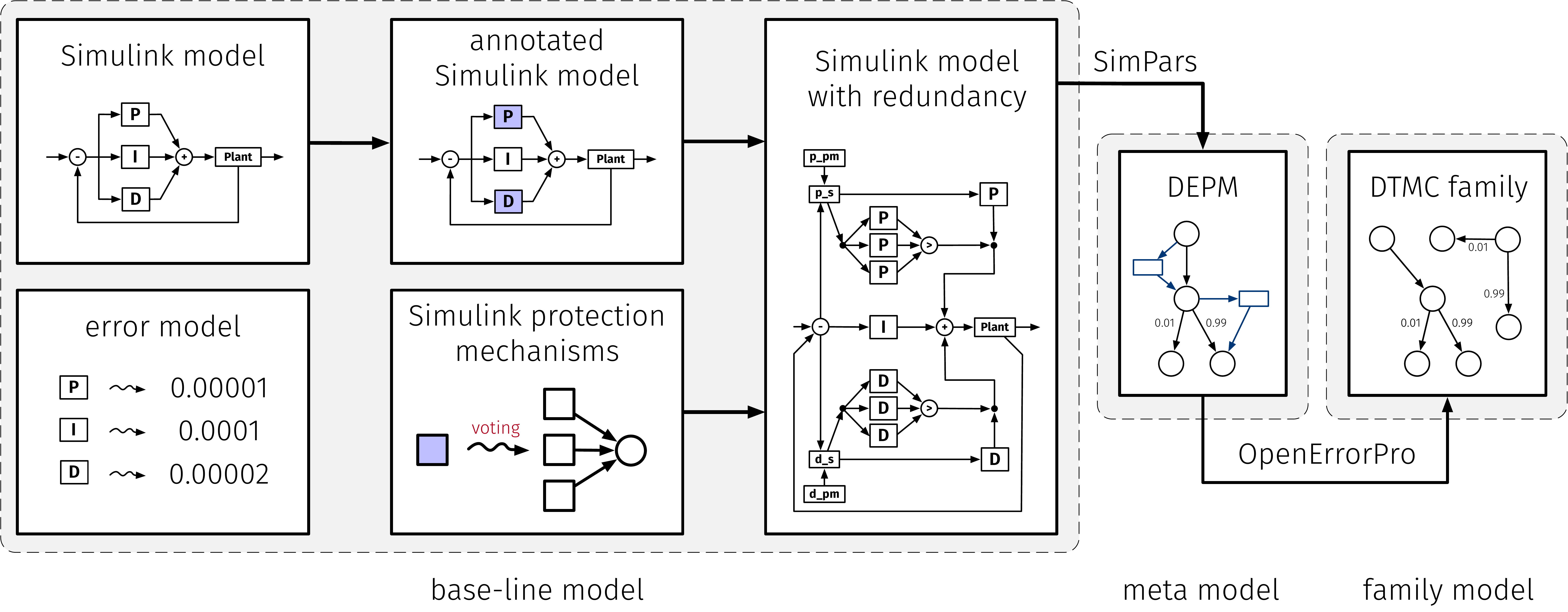}
	\caption{\label{fig:schema} Schema of the approach to obtain DTMC 
		families from annotated $\simulink$ models~\cite{DDMBJ19}}
\end{figure*}

\noindent For $\simulink$ models with protection mechanisms, we introduced 
a discrete Markov chain (DTMC) family semantics~\cite{DDMBJ19}
that can be automatically generated using $\errorpro$~\cite{MorDinSte19a}. 
The general workflow of our approach towards 
a DTMC family out from $\simulink$ models is 
depicted in Figure~\ref{fig:schema}.

\noindent First, blocks in the $\simulink$ model are annotated with the types 
of protections that should be considered, e.g., comparison or voting.
In Figure~\ref{fig:schema}, we annotated voting protections for the 
P and D block (indicated by shaded blocks). 
Then, the annotated model is transformed to
a $\simulink$ model that includes redundancies by replacing every block
where protection should be taken into account with a switch 
block that depends on a fresh switch variable, followed by 
the actual blocks for the redundancy mechanisms. 
The role of the switch variable is to select the redundancy 
mechanism, directing the control flow, e.g., to the triplicated
blocks in case of the voting mechanism.
The $\simulink$ model with redundancy stands for a
family of models with different protection combinations.
An error model furthermore assigns to each $\simulink$ block 
the probability for some fault occurring in this block,
introducing stochastics into the model required for a
reliability analysis.
In~\cite{DDMBJ19} we presented a DTMC family semantics
as $\prism$ family model generated using the tools 
$\simpars$~\cite{MorJanKru16a} and $\errorpro$~\cite{MorDinSte19a}.
For this, we employed a meta-model representation as
Dual-graph Error Propagation Model (DEPM)~\cite{MorDinSte19a}
(see also Figure~\ref{fig:schema}).

\subsection{The Velocity Control Loop Model}
We illustrate our approach of iterative variable reordering described
in Section~\ref{sec:reordering} on a $\prism$ program that stems from an aircraft 
velocity control loop (VCL) with redundancies described in~\cite{DDMBJ19}. 
Our model is a simplified version of the aircraft model borrowed 
from the $\simulink$ example set~\cite{aircraft} that itself is based on 
a long-haul passenger aircraft flying at cruising altitude and speed, 
adjusting the fuel flow rate to control the aircraft velocity.
Figure~\ref{fig:velocity} shows the  $\simulink$ model where all 
blocks are amendable for protection mechanisms.
In~\cite{DDMBJ19} we applied three different kinds of redundancy mechanisms 
on eight of these blocks, leading a $\prism$ DTMC family model with 
$4^8=65\,536$ family members, each member standing for one protection 
combination. 
The $\prism$ family model generated by $\errorpro$~\cite{MorDinSte19a}
could not immediately be constructed by $\prism$'s symbolic engine.
We hence had to apply two (handcrafted) optimizations to enable a reliability
analysis of the VCL redundancy system: \emph{reset value optimization}~\cite{GarSer06a,DubMorBai20}
and \emph{iterative variable reordering}. While reset value optimization speeded up 
the analysis time, iterative variable reordering was the technique 
that enabled the construction of the model.

In this paper, we use a variant of the VCL model of \cite{DDMBJ19}
as starting point where neither reset value optimization nor hand-crafted iterative 
variable reordering has been applied to and where 13 blocks could be
protected with either voting or comparison mechanisms.

\section{Implementation and Evaluation}\label{sec:evaluation}
In this section, we report on an implementation of the iterative 
variable reordering for family models expressed in the input
language of $\prism$ that follows Algorithm~\ref{ireorder-pseudo} 
and includes the optimization heuristics detailed in 
Section~\ref{sec:reordering}.

\subsection{Implementation}
Our implementation is done in $\python$ and takes as input a $\prism$
family model, a parameter specifying the variable selection heuristics
($\pi$-minimal, $\rho$-minimal, or $\rho$-maximal), and the step size
of the iteration.
The advantage of the heuristics is that the construction of the
model can be parallelized trying different heuristics on
multi-core systems to eventually obtain a suitable variable
order that renders the whole system family to be constructible even
when some heuristics are slow or not successful.

\paragraph{Remark.} 
$\prism$ provides a faster model construction for non-family models,
i.e., models without an \texttt{init} block, than for family models as evaluating
the \texttt{init}-block expression relies on SAT solving, an NP-complete problem~\cite{GJ79}. 
Hence, we implemented the initial construction step
in line 1 of Algorithm~\ref{ireorder-pseudo} for a model representation 
without \texttt{init} block. For later constructions (line 9 of
Algorithm~\ref{ireorder-pseudo}) generated an \texttt{init} block
for $\iota\wedge\cnf(E)$ as indicated.

\subsection{Evaluation}
We evaluated our implementation on the $\prism$ DTMC family generated
from a $\simulink$ VCL redundancy model (see Section~\ref{sec:model})
where 13 blocks are marked as protectable by comparison or voting mechanisms.
Thus, we applied iterative variable reordering on a family model with
$3^{13}=1\,594\,323$ protection variants, more than 24 times
the number of protections considered in~\cite{DDMBJ19}.
Note that we did not apply reset value 
optimization~\cite{GarSer06a,DubMorBai20} as in~\cite{DDMBJ19}.
All experiments are carried out on a Linux server system\footnote{2 $\times$ Intel Xeon 
E5-2680 (Octa Core, Sandy Bridge) running at 2.70~GHz with 384~GB of RAM; Turbo Boost 
and Hyper Threading disabled; Debian GNU/Linux 9.1.} with a memory bound
of 30~GB of RAM, running the $\prism$ version presented in~\cite{KBCDDKMM18}.

\paragraph{Heuristic Comparison.}
First, we evaluate the different heuristics implemented.
Table~\ref{tab:heuristics} shows the situation of the implementation
as snapshot after 20 minutes.
\begin{table}
	\caption{\label{tab:heuristics} Performance of the heuristics within 20 minutes}
	\tt
	\begin{center}
		\begin{tabular}{l|r||r|r|r|r}
			variable selection& step size& iteration & combinations & states & nodes\\\hline
			$\pi$-minimal  	&1& 22& 177\,147 & 8.67$\cdot 10^{11}$ & 97\,386\\
			\rowcolor{blue!10}&2& 12& 531\,441 & 6.19$\cdot 10^{12}$ & 130\,321\\
							&3& 7& 118\,098 & 6.29$\cdot 10^{11}$ & 101\,766\\
							&4& 5& 59\,049 & 1.25$\cdot 10^{11}$ & 88\,054\\\hline
			$\rho$-minimal 	&1& 22& 177\,147 & 8.67$\cdot 10^{11}$ & 97\,386\\
							&2& 12& 531\,441 & 7.37$\cdot 10^{12}$ & 153\,189\\
							&3& 7& 118\,098 & 8.17$\cdot 10^{11}$ & 138\,287\\
							&4& 5& 59\,049 & 1.19$\cdot 10^{11}$ & 163\,882\\\hline
			$\rho$-maximal 	&1& 12& 4\,096 & 9.47$\cdot 10^{10}$ & 126\,338\\
							&2& 10& 59\,049 & 6.83$\cdot 10^{11}$ & 139\,991\\
							&3& 5& 7\,776 & 1.25$\cdot 10^{11}$ & 251\,944\\
							&4& 5& 59\,049 & 9.71$\cdot 10^{11}$ & 225\,711
		\end{tabular}

	\end{center}
\end{table}
As expected, the higher the step size, the less iterations could be performed
by the iterative variable reordering. 
On our model, $\pi$-minimal and $\rho$-minimal selection heuristics perform
best with step size 2. For both heuristics, the complete family 
model with its $3^{13}=1\,594\,323$ 
protection variants was constructed within half an hour.
It is not surprising that step size 2 is a good
choice as we considered two redundancy mechanisms for each block, i.e.,
in each step all redundancies for each block contribute
to the family model at once.

\paragraph{Iteration Statistics.}
For step size of 2 and using $\pi$-minimal variable selection,
we detail the statistics of every iteration
until the whole family model has been constructed in Table~\ref{tab:whole}.
We shaded the iteration that was reached after 45 minutes
of computation time (cf. the shaded row in Table~\ref{tab:heuristics}).
\begin{table}

	\caption{\label{tab:whole} Step-wise statistics of $\rho$-maximal }
\tt 
	\begin{center}
		\begin{tabular}{r|r|r|r|r|r|r}
		& & &\multicolumn{2}{c|}{\#nodes} & \multicolumn{2}{c}{time [s]}\\
iteration & combinations & states & before & after & model & reorder \\\hline
0	&1			&113\,891			&24\,816	&23\,359	&7.40	&2.67\\
1	&3			&347\,401			&24\,576	&23\,329	&7.81	&2.87\\
2	&9			&1.06$\cdot 10^{6}$	&24\,545	&23\,319	&7.36	&2.93\\
3	&27			&3.93$\cdot 10^{6}$	&48\,413	&33\,647	&8.48	&4.76\\
4	&81			&1.19$\cdot 10^{7}$	&35\,185	&33\,664	&8.30	&4.55\\
5	&243		&4.05$\cdot 10^{7}$	&52\,497	&40\,431	&9.18	&6.12\\
6	&729		&1.54$\cdot 10^{8}$	&69\,535	&48\,888	&13.76	&8.07\\
7	&2\,187		&6.65$\cdot 10^{8}$	&94\,022	&55\,835	&23.57	&11.00\\
8	&6\,561		&3.00$\cdot 10^{9}$	&73\,242	&62\,636	&28.55	&9.75\\
9	&19\,683	&1.85$\cdot 10^{11}$&132\,279	&76\,518	&73.55	&18.14\\
10	&59\,049	&1.25$\cdot 10^{11}$&95\,740	&79\,420	&76.40	&14.63\\
11	&177\,147	&8.67$\cdot 10^{12}$&120\,565	&97\,386	&155.03	&19.08\\\rowcolor{blue!10}
12	&531\,441	&6.19$\cdot 10^{12}$&205\,962	&130\,321	&360.38	&33.37\\\hline
13	&1\,594\,323&4.87$\cdot 10^{13}$&256\,662	&160\,798	&723.02	&41.08\\\hline
$\sum$ & \multicolumn{3}{c}{} &&1\,502.80 & 179.00
		\end{tabular}
	\end{center}
\end{table}
In the column \texttt{combinations}, the family size of the intermediate
family is listed. Column \texttt{\#nodes} shows 
the number of MTBDD-nodes used to represent that family model
before applying reordering (cf. line 10 of Algorithm~\ref{ireorder-pseudo})
and thereafter. The last row indicates the time required to construct the model
(cf. line 9 of Algorithm~\ref{ireorder-pseudo}) and to perform
reordering (cf. line 10 of Algorithm~\ref{ireorder-pseudo}).

\paragraph{Alternative Approaches.}
Clearly, there are alternative methods that could be used to enable
an analysis of large-scale system families when the family model
cannot be constructed in an ad-hoc fashion. 
The first is to switch from an all-in-one
approach to a one-by-one approach, i.e., analyzing every family
member in isolation. However, even the smallest family member
(see Table~\ref{tab:whole}, first iteration) required more than 7 seconds
for the construction of the model. One could hence expect at least 
$3^{13}\cdot 7$ seconds of construction time for all family members,
which would correspond to more than one year of construction time on
our test system. 
Second, one could apply dynamic reordering techniques as implemented
in the probabilistic model checker $\storm$~\cite{DJKV-CAV17}.
With this method, the variable order on the MTBDD is dynamically
changed during the model-construction process. However,
even with smaller families, e.g., the VCL family model from~\cite{DDMBJ19}
where only 8 protectable blocks were considered, 
model construction did not finish within days.
We hence conclude that, at least for the VCL family models we generated,
iterative variable reordering is the only technique we considered and 
yielded a successful model construction.

\section{Concluding Remarks}\label{sec:conclusion}
In this paper we presented iterative variable reordering as a generic method
to cope with the problem of constructing family models of huge system families.
While our implementation supports models specified in the input language
of the probabilistic model checker $\prism$, the formally defined algorithm
is applicable to a wide range of analysis tools that use variable-order sensitive
symbolic representations for their models and support some kind of reordering
mechanism. For instance, the verification of large feature-oriented system families 
according to \cite{ClassenHSL11} those verification is based on the BDD engine of 
$\nusmv$ also could benefit from our approach.
By supporting system families given as $\prism$ programs, 
our implementation can be directly applied on feature-oriented systems 
specified in $\profeat$~\cite{CDKB18}. Our methods could be also viable
for family-based parameter synthesis~\cite{BD18}.
Experiments on such system families are left for future work.

\paragraph{Acknowledgements.}
We would like to thank Joachim Klein who extended $\storm$
with support for family models, used to validate that dynamic reordering
techniques are not viable for the VCL example. 
\bibliographystyle{eptcs}
\bibliography{main}
\end{document}